\documentclass[]{spie}  %>>> use for US letter paper
\usepackage{amsmath,amsfonts,amssymb}
\usepackage{array,multirow,graphicx}
\usepackage{adjustbox}
\usepackage{booktabs}
\usepackage{tabu, colortbl}
\usepackage{xr}
\usepackage[dvipsnames]{xcolor}
\usepackage[super]{nth}
\usepackage{tikz}
\usepackage[colorlinks=true, allcolors=blue]{hyperref}

\definecolor{mycorrect}{rgb}{1, 0, 0} % red
\definecolor{light_grey}{rgb}{0.6, 0.6, 0.6}

\title{Pre- to Post-Contrast Breast MRI Synthesis for Enhanced Tumour Segmentation}

\author[a, b, c]{Richard Osuala}
\author[a]{Smriti Joshi}
\author[d]{Apostolia Tsirikoglou}
\author[a]{Lidia Garrucho}
\author[e]{Walter H. L. Pinaya}
\author[a]{Oliver Diaz}
\author[a]{Karim Lekadir}
\affil[a]{%Barcelona Artificial Intelligence in Medicine Lab (BCN-AIM), 
Departament de Matemàtiques i Informàtica, Universitat de Barcelona, Spain}
\affil[b]{Helmholtz Center Munich, Munich, Germany}
\affil[c]{Technical University of Munich, Munich, Germany}
\affil[d]{Karolinska Institutet, Sweden}
\affil[e]{King’s College London, London, United Kingdom}

\authorinfo{
Corresponding author: Richard Osuala (richard.osuala@ub.edu)}

\begin{document} 

\maketitle

\begin{abstract}

Despite its benefits for tumour detection and treatment, the administration of contrast agents in dynamic contrast-enhanced MRI (DCE-MRI) is associated with a range of issues, including their invasiveness, bioaccumulation, and a risk of nephrogenic systemic fibrosis. This study explores the feasibility of producing synthetic contrast enhancements by translating pre-contrast T1-weighted fat-saturated  breast MRI to their corresponding first DCE-MRI sequence leveraging the capabilities of a generative adversarial network (GAN). Additionally, we introduce a Scaled Aggregate Measure (SAMe) designed for quantitatively evaluating the quality of synthetic data in a principled manner and serving as a basis for selecting the optimal generative model.  We assess the generated DCE-MRI data using quantitative image quality metrics and apply them to the downstream task of 3D breast tumour segmentation. Our results highlight the potential of post-contrast DCE-MRI synthesis in enhancing the robustness of breast tumour segmentation models via data augmentation. Our code is available at \url{https://github.com/RichardObi/pre_post_synthesis}.

%%% A longer detailed version of the abstract
%Despite its benefits for tumour detection and treatment, the administration of contrast agents in dynamic contrast-enhanced MRI (DCE-MRI) is associated with a range of issues, including their invasiveness, bioaccumulation, and a risk of nephrogenic systemic fibrosis. This study explores the feasibility of producing synthetic contrast enhancements by translating pre-contrast T1-weighted breast MRI to their corresponding first DCE-MRI sequence leveraging the capabilities of a Pix2PixHD generative adversarial network (GAN). Additionally, we introduce a Scaled Aggregate Measure (SAMe) designed for quantitatively evaluating the quality of synthetic data in a principled manner based on a scaled combination of perceptual domain-specific and domain-agnostic Fréchet Inception Distances (FID), and pixel-level metrics, namely, mean squared error (MSE), mean absolute error (MAE) and structural similarity index measure (SSIM).  We apply SAMe to serve as a basis for selecting the optimal generative model across training epochs. We further assess the generated DCE-MRI data using the aforementioned quantitative image quality metrics and showcase its advantages in the downstream task of 3D breast tumour segmentation based on a 3D U-Net model. Our results highlight the potential of post-contrast DCE-MRI synthesis in enhancing the performance and robustness of breast tumour segmentation models via data augmentation.

\end{abstract}
% Include a list of keywords after the abstract 

\section{INTRODUCTION}
\label{sec:intro}  % \label{} allows reference to this section
% 1. Defining the problem extension of breast cancer.
In 2020, breast cancer stood out as the most prevalent cancer type worldwide across all age groups and genders. 
With a staggering 2.26 million new cases and 684,996 deaths reported, breast cancer's global impact is profound \cite{globalCancerObservatory}. %rephrased 06082023
%%%%%%% Improved treatment with DCE-MRI and contrast agents.
Amid detection methods, breast dynamic contrast-enhanced magnetic resonance imaging (DCE-MRI) emerges as remarkably sensitive compared to alternatives such as mammography or ultrasound \cite{mann2019contrast}. 
Beyond screening, DCE-MRI finds widespread use in breast cancer diagnosis and treatment, serving vital roles in 
monitoring, preoperative planning, treatment and neoadjuvant therapy response assessment, where the radiological response is asssessed through lesion size regress/progress \cite{ground_truth, radhakrishna2018role}.
%
%%%%%%% Dangers and disadvantages of contrast agents
However, the administration of gadolinium-based contrast agents is associated with a range of adverse risks and side effects. These include the deposition of residual substances and their bioaccumulation with unclear clinical significance and long-term consequences 
\cite{idee2006clinical, marckmann2006nephrogenic,kanda2014high, nguyen2020dentate, olchowy2017presence}, 
as well as the increased potential for nephrogenic systemic fibrosis \cite{marckmann2006nephrogenic}. After a review of gadolinium-based contrast agents requested by the European Commission in 2016, the European Medicines Agency recommended restrictions for some intravenous linear agents to prevent risks potentially associated with gadolinium deposition \cite{euagency2017}.
Beyond these concerns, the very process of gadolinium-based contrast agent administration is replete with drawbacks, including time-consuming protocol,  substantial financial outlays, susceptibility to allergic reactions, and the necessity of intravenous cannulation coupled with the cumbersome injection of the contrast media. These collective factors converge to impose an unwarranted burden upon the patient, encompassing dimensions of inconvenience and potential compromise of well-being \cite{zhang2023synthesis, olchowy2017presence, nguyen2020dentate}. 

%%%%%%% Desire to minimise use of contrast agents
With recent advances in deep learning, training deep generative models to generate synthetic contrast-enhanced imaging data as an alternative to contrast agent administration 
has been becoming a promising field of research 
\cite{osuala2022data, pasquini2022synthetic}.
For instance, Kim et al. \cite{kim2022tumor} provide a tumour-attentive segmentation-guided generative adversarial network (GAN) \cite{goodfellow2014generative} that generates a contrast-enhanced T1 breast MRI image from its pre-contrast counterpart, while being guided by the predictions of a surrogate segmentation network. %With one objective being to improve segmentation using the GAN-generated data, it can become counterproductive to limit the GAN contrast translation to the tumour segmentation predicted by the segmentation model. 
Similarly, Zhao et al. \cite{zhao2020tripartite} propose Tripartite-GAN to synthesise contrast-enhanced from non contrast-enhanced liver MRI with a chained tumour detection model.
However, high-quality annotations such as segmentation masks may be scarce and therefore it is desirable to achieve pre- to post-contrast translation without relying on annotations.
Xue et al. \cite{xue2022bi} propose a bi-directional pre- to post-contrast and post- to pre-contrast brain MRI image translation network based on pix2pix \cite{isola2017image} with contrast and image encoded in separate latent representations. 
%Their encoder encodes contrast and image separately with the contrast representation producing a contrast enhancement map as output that can be subtracted from the (generated) post-contrast image to get the pre-contrast image. 
Wang et al. \cite{wang2021synthesizing} propose a two-stage GAN, where in the first stage the contrast enhancement of the T1-weighted image is segmented based on an adversarial loss. In the second stage, a synthetic post-contrast DCE image generator is trained, which depends on the accuracy of the segmentation network from the first stage.
%with is trained based on (a) a L1-loss, (b) an adversarial loss, and (c) an edge detector based L2 loss.
Müller-Franzes et al. \cite{muller2023using} translate T1 and T2 images to post-contrast breast MRI images using a pix2pixHD \cite{wang2018high} to test image realism in a reader study. 
Han et al. \cite{han2023synthesis} model the translation of Diffusion Weighted Imaging (DWI) from DCE breast MRI volumes as sequence-to-sequence translation task, while Zhang et al. \cite{zhang2023synthesis} design a GAN to synthesise contrast-enhanced breast MRI from a combination of encoded T1-weighted MRI and DWI images.
However, such recent promising approaches \cite{wang2021synthesizing, xue2022bi, muller2023using, zhang2023synthesis} 
are not validated %for the clinically relevant task of 
on their potential to improve tumour segmentation using synthetic data. % i.e. whether the synthetic images can enhance tumour segmentation models.

The surveyed studies use multiple different image quality evaluation metrics to evaluate image synthesis. The indicated absence of a consensus metric 
 motivates our proposal of a unified synthetic data quality measure. 
%%%%% Contributions
In summary, the main contributions of our work are:\vspace{-0.13cm} %  as follows
\begin{itemize}
    \item A GAN-based synthesis model to effectively translate pre- to post-contrast breast MRI axial slices. 
     \vspace{-0.13cm} 
    \item Proposing the Scaled Aggregate Measure (SAMe), combining perceptual and pixel-level metrics for principled generative model comparison and training checkpoint selection.
    \vspace{-0.13cm} 
    \item Demonstrating the potential of synthetic DCE-MRI to enhance breast tumour segmentation robustness.
    %\vspace{-0.13cm} 
\end{itemize}

% Methods
\section{MATERIALS AND METHODS}
\label{sec:methodology}

\subsection{Dataset}
The dataset used in the present study is the single-institutional open-access Duke-Breast-Cancer-MRI Dataset \cite{saha2018machine}. The data was gathered in Duke Hospital (US) in the timeframe from \nth{1} January 2000 to \nth{23} March 2014 and consists of 922 biopsy-confirmed patient cases with invasive breast cancer and available pre-operative MRI at Duke Hospital.
The MRI voxel dimensions are either $448 \times 448$ or $512 \times 512$ in the coronal and sagittal planes, while the number of slices in the axial plane is variable. 
The MRI images are acquired using a magnetic field strength of 1.5 T or 3 T. 
Each breast cancer case comprises one fat-saturated T1 sequence alongside up to 4 corresponding DCE fat-saturated T1-weighted sequences.
%alongside corresponding non-imaging data such as demographic and treatment data, as well as MRI findings, tumour characteristics, and recurrence and follow-up information.% with contours annotations of tumour locations. 
%The dataset contains an MRI volume of the pre-contrast sequence as well as several 
These T1-weighted DCE post-contrast sequences are acquired after contrast agent injection, with a median of 131 seconds passed between post-contrast acquisitions. 
%In this study, we translate from the fat-saturated T1 pre-contrast sequence to the first phase of the fat-saturated T1-DCE sequences in axial orientation.
%The tumour contours are available for all of the 922 patients/cases and are defined by column and row pixel information alongside the start and end slices in the 3D MRI volume, which allows to extract 3D and 2D region-of-interest bounding boxes.
%This means that each patient was scanned several times e.g., with a few seconds passing between (once per phase) and that there might be slight differences in patient positioning for each phase. The dataset contains a multitude of non-imaging patient data such as:
%\begin{enumerate}
%    \item Demographic data (date of birth, menopause, ethnicity, metastasis information).
%    \item Tumour characteristics data (e.g., ER, PR, HER2, tumour grade).
%    \item Treatment data (e.g., types of treatments, surgery, chemotherapy, radiation therapy, tumour response, neoadjuvant therapy).
%    \item MRI findings data (Multicentric/Multifocal, Contralateral Breast Involvement, Lymphadenopathy or Suspicious Nodes Skin/Nipple Invovlement, Pec/Chest Involvement).
%    \item Recurrence and follow-up (e.g., recurrence events, days to recurrence).
%    \item Data from corresponding mammography (age at mammo, breast density, tumour shape, margin, architectural distortion, mass density, calcifications, tumor size).
%    \item Ultrasound features (e.g., shape, margin, tumour size, echogenicity).
%\end{enumerate}
3D tumour segmentation masks are provided for 254 cases from the authors of a related work \cite{ground_truth}. These annotations were 
automatically segmented by a fuzzy means algorithm in MATLAB, revised by an experienced medical physicist, and verified by a radiologist. 
%We manually verified these 254 cases validating that the masks correctly correspond to the tumour volumes in the T1 DCE phase 1 sequence. %251 out of these 254 cases further include a pathological treatment response (PCR) label. Out of the total of 922, the PCR label is available for 300 cases of which 251 have annotated segmentation masks.
For the image synthesis model, 668 cases without segmentation masks, out of the total of 922 cases of the dataset, are used as training data, while the remaining 254 cases with masks are randomly split into validation (224 cases) and test (30 cases) sets.
For the segmentation model, the same test set is used, while for training and validation 33 multi-focal cases are removed from the cases with masks before applying a 5-fold cross-validation, splitting the remaining 191 cases into training (80\%) and validation (20\%) subsets. 

\subsection{Pre- to Post-Contrast DCE-MRI Synthesis}

\begin{figure} [ht]
\begin{center}
\includegraphics[height=7.2cm]{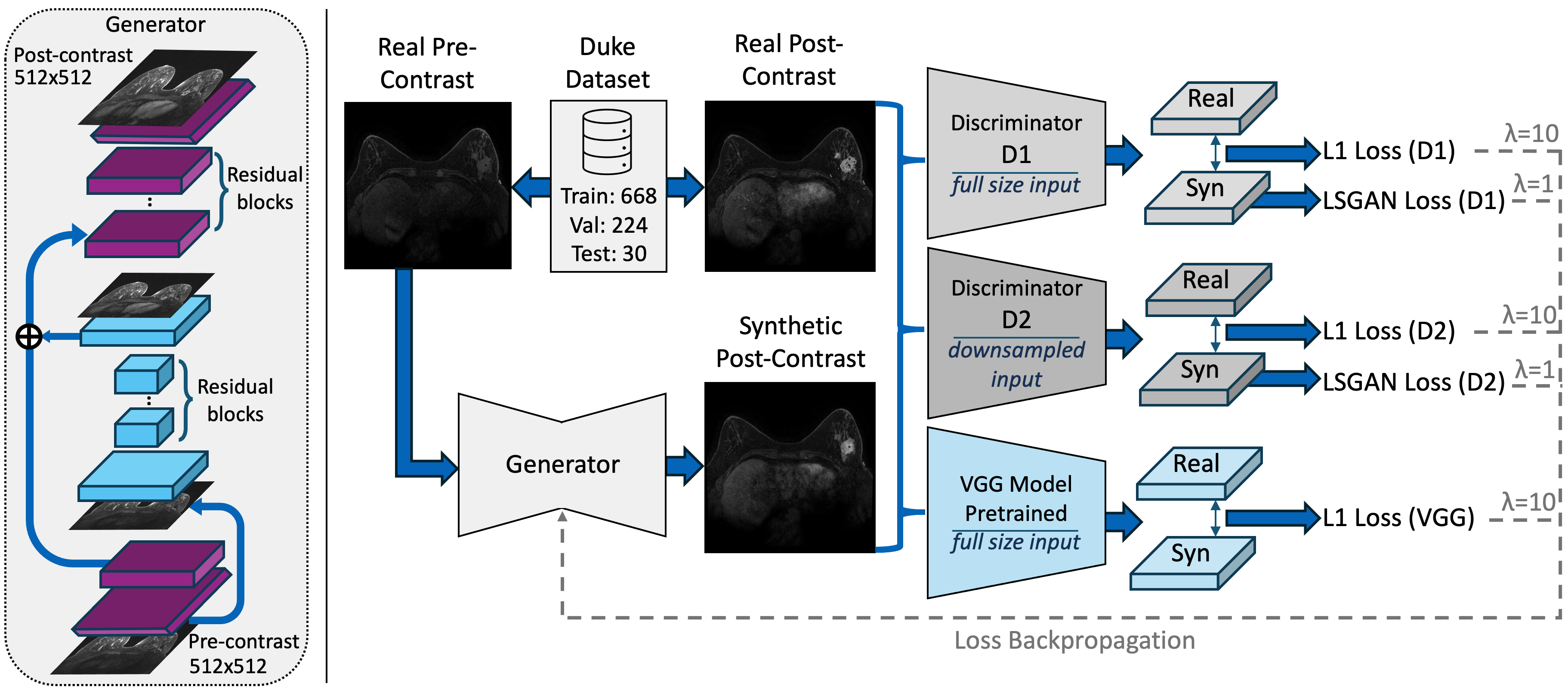}
\end{center}
%\vspace{-6mm}
\caption[] 
{\label{fig:gan_training} Overview of training workflow of our pre- to post-contrast translating Generative Adversarial Network (GAN) based on Pix2PixHD \cite{wang2018high}. Three reconstruction losses (L1) and two least squares adversarial losses (LSGAN)\cite{mao2017least} from two discriminators (D1 \& D2) and one pretrained VGG \cite{simonyan2014very} model are backpropagated into the generator, where lambda ($\lambda$) represents the weight of each of the different losses. Processing the images at two different scales inside the generator architecture balances local detail and global consistency, which is further enforced by the two different image input scales in D1 (full size) and D2 (downsampled).
}
\end{figure}

GANs \cite{goodfellow2014generative} are based on a two-player min-max game of a generator and a discriminator network. The generator ($G$) strives to create samples ($\hat{x}$) from a noise distribution ($p_{z}$) that the discriminator ($D$) cannot distinguish from samples ($x$) from the real image distribution ($p_{data}$), resulting in the value function of Equation \ref{eq:1}.
\begin{equation} \label{eq:1}
\begin{aligned}
\min_{G} \max_{D} V(D,G) = \min_{G} \max_{D}[\mathbb{E}_{x\sim p_{data}} [log D(x)] + \mathbb{E}_{z\sim p_{z}} [log(1 - D(G(z)))]].
\end{aligned}
\end{equation}
In the image-to-image (I2I) translation scenario, instead of sampling from a noise distribution, GANs \cite{goodfellow2014generative} are given an input sample from a source distribution ($x$) to synthesise a corresponding sample from a target distribution ($\hat{y}$). In this work, we adopt a Pix2PixHD\cite{wang2018high} GAN for I2I translation from pre- to post-contrast images resized and stacked to $512 \times 512 \times 3$ pixel dimensions. Pix2PixHD is chosen due to its capabilities of generating high-quality cancer imaging data \cite{osuala2022data} and due to its network architecture and methodological setup specifically designed for paired image-to-image (i.e., pre-to-post-contrast) translation. %without the need to generate unobservable counterfactuals in unpaired image-to-image translation scenarios. 
As shown in figure \ref{fig:gan_training}, the GAN consists of a generator network processing two image scales, one enforcing global consistency and the other for the generation of finer details. It further includes two identical discriminator networks that also operate at different image scales based on the downsampling of the input images. The model is trained using a weighted ($\lambda$) combination of least squares adversarial losses \cite{mao2017least} ($\lambda$=1), discriminator feature matching losses ($\lambda=10$) implemented as summed L1-loss between the extracted real and fake image features of each of the two discriminators, as well as a VGG-based \cite{simonyan2014very} perceptual loss ($\lambda=10$). Input images are transformed into the range $[-1, 1]$ and have a probability of 50\% of being rotated by 90 degrees during training. The model is trained for 200 epochs using an Adam optimiser ($\beta=0.5$) and a learning rate of 2e-4 that decays linearly to zero from epoch 100 to 200. The 2D grayscale input images are stacked in 3-channels and resized to $512 \times 512$ pixel dimensions. The model is trained on a NVIDIA GeForce RTX 3090 GPU with 24GB RAM with a batch size of 1.
Before model training, 2D PNG images are extracted from the fat-saturated T1 MRI (source domain input) and T1 DCE phase 1 NIfTI volumes (target domain output) and resized to maintain an aspect ratio of (1,1,1). The slices are extracted in axial dimension from a wide range starting at slice 1 up to slice 196 to allow the model to learn to translate any slice of the 3D MRI volume rather than only slices containing tumours. For validation and test data, each 2D slice is translated iteratively from pre- to post-contrast to assemble 3D synthetic post-contrast volumes.

\subsection{Tumour Segmentation}
\begin{figure} [ht]
\begin{center}
\includegraphics[height=5.5cm]{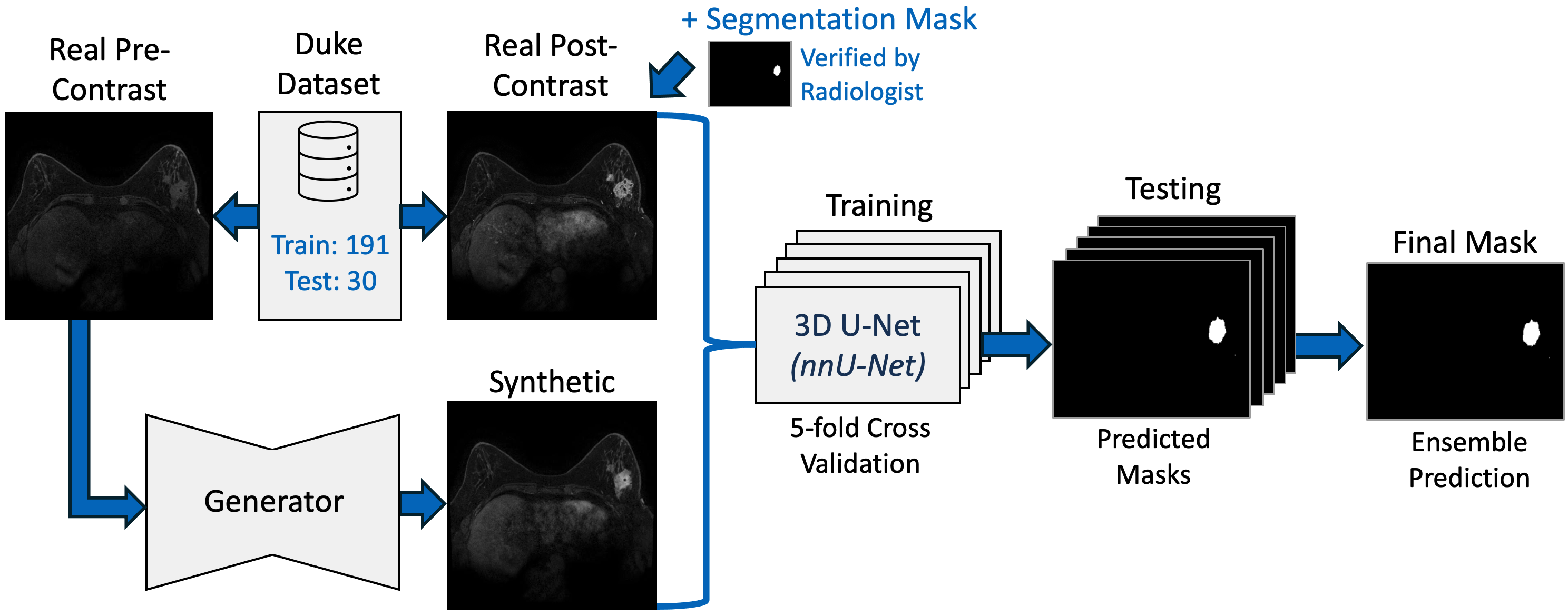}
\end{center}
%\vspace{-6mm}
\caption[] 
{\label{fig:segmentation} Overview of the segmentation method based on 3D U-Nets \cite{ronneberger2015u} from the nnU-Net \cite{isensee2021nnu} framework. The iteratively translated synthetic post-contrast axial slices are stacked to create 3D breast MRI volumes. These synthetic volumes correspond to the tumour segmentation masks, which were initially acquired based on the real post-contrast fat-saturated sequence.
}
\end{figure}

\begin{figure}
    \centering
    \begin{tikzpicture}
    \draw (0, 0) node[inner sep=0] {
        \includegraphics[height=2.25cm, width=.164\textwidth, trim={0 4.5cm 0 2.5cm},clip]{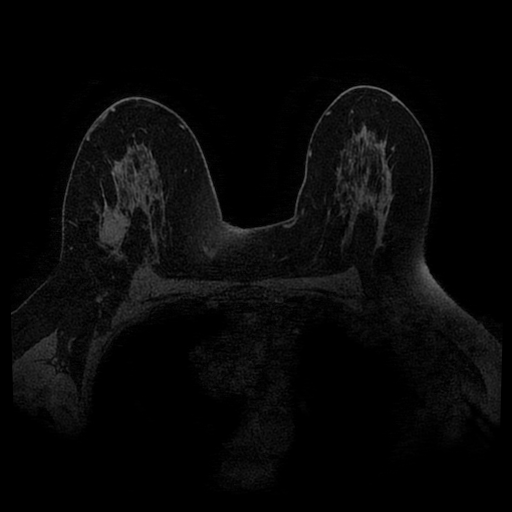}%\hfill
            };
    \draw (0, 1.4) node {\tiny(\emph{a}) pre-contrast};
    \end{tikzpicture}\hspace{0cm}%
    \begin{tikzpicture}
    \draw (0, 0) node[inner sep=0] {
    \includegraphics[height=2.25cm, width=.164\textwidth, trim={0 4.5cm 0 2.5cm},clip]{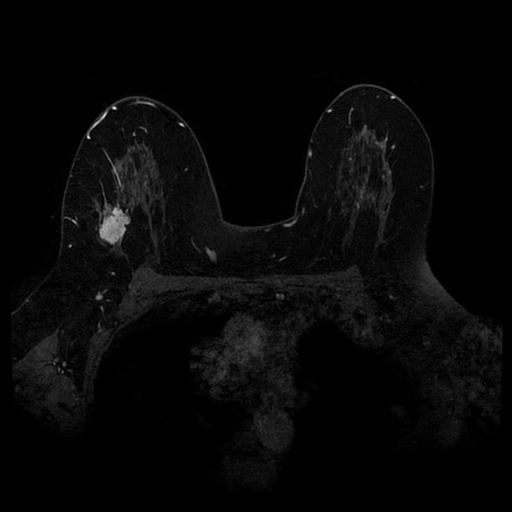}%\hfill
        };
    \draw (0, 1.4) node {\tiny(\emph{b}) real post-contrast};
    \end{tikzpicture}\hspace{0cm}%
    \begin{tikzpicture}
    \draw (0, 0) node[inner sep=0] {
    \includegraphics[height=2.25cm, width=.164\textwidth, trim={0 4.5cm 0 2.5cm},clip]{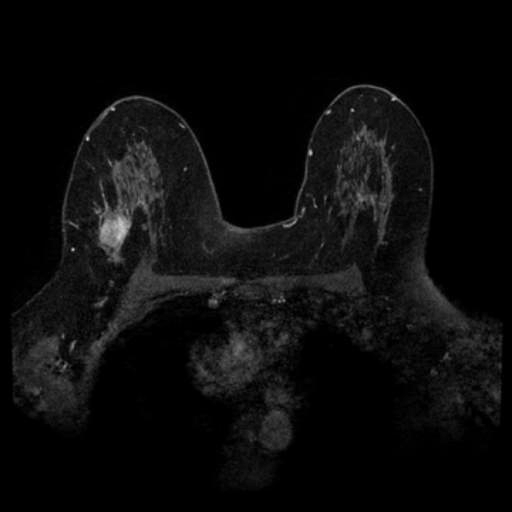}%\hfill
    };
    \draw (0, 1.4) node {\tiny(\emph{c}) syn post-contrast};
    \end{tikzpicture}\hspace{0cm}%
    \begin{tikzpicture}
    \draw (0, 0) node[inner sep=0] {
    \includegraphics[height=2.25cm, width=.164\textwidth, trim={0 4.5cm 0 2.5cm},clip]{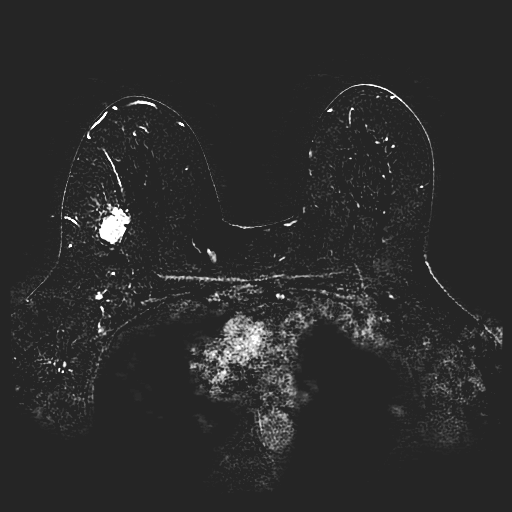}%\hfill
    };
    \draw (0, 1.4) node {\tiny(\emph{d}) real subtraction};
    \end{tikzpicture}\hspace{0cm}%
    \begin{tikzpicture}
    \draw (0, 0) node[inner sep=0] {
    \includegraphics[height=2.25cm, width=.164\textwidth, trim={0 4.5cm 0 2.5cm},clip]{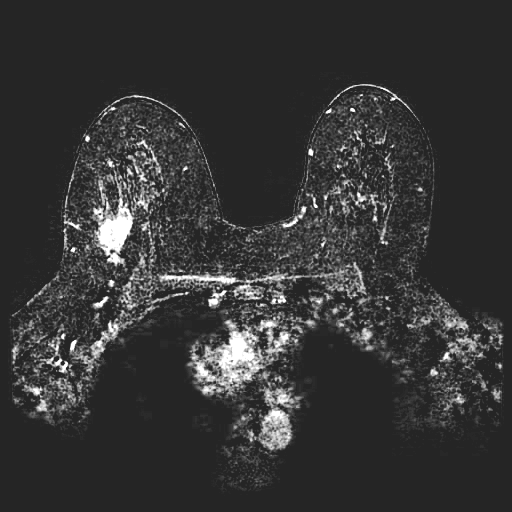}%\hfill
    };
    \draw (0, 1.4) node {\tiny(\emph{e}) syn subtraction};
    \end{tikzpicture}\hspace{0cm}%
    \begin{tikzpicture}
    \draw (0, 0) node[inner sep=0] {
    \includegraphics[height=2.25cm, width=.164\textwidth, trim={0 4.5cm 0 2.5cm},clip]{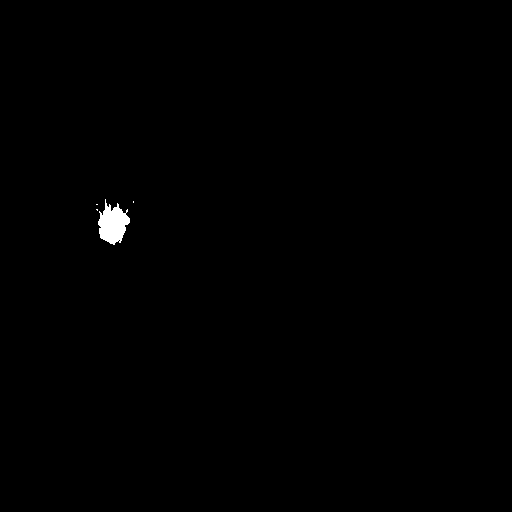}%\vspace{0.1cm}
    %\\[\smallskipamount]
    };
    \draw (0, 1.4) node {\tiny(\emph{f}) GT mask};
    \end{tikzpicture}\hspace{0cm}%

    \includegraphics[height=2.25cm, width=.164\textwidth, trim={2.cm 0cm 0.5cm 0},clip]{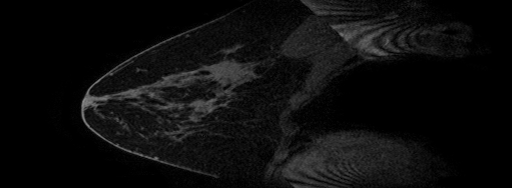}%\hfill
    \includegraphics[height=2.25cm, width=.164\textwidth, trim={2.cm 0cm 0.5cm 0},clip]{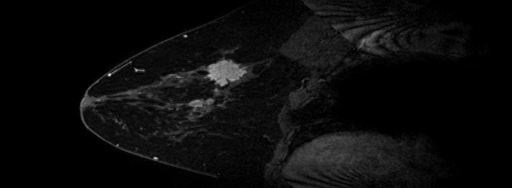}%\hfill
    \includegraphics[height=2.25cm, width=.164\textwidth, trim={2.cm 0cm 0.5cm 0},clip]{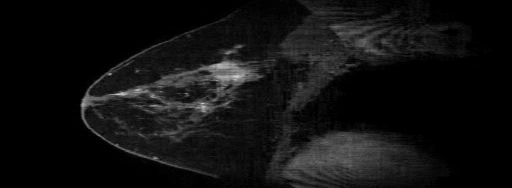}%\hfill
    \includegraphics[height=2.25cm, width=.164\textwidth, trim={2.cm 0cm 0.5cm 0},clip]{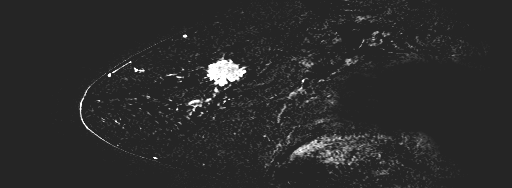}%\hfill
    \includegraphics[height=2.25cm, width=.164\textwidth, trim={2.cm 0cm 0.5cm 0},clip]{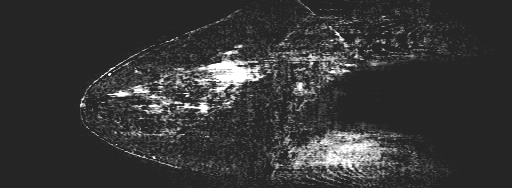}%\hfill
    \includegraphics[height=2.25cm, width=.164\textwidth, trim={2.cm 0cm 0.5cm 0},clip]{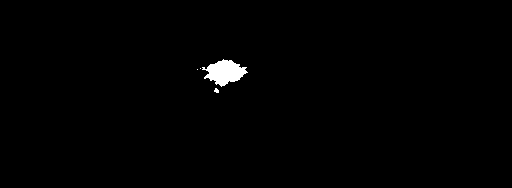}
    %\\[\smallskipamount]

    \includegraphics[height=2.25cm, width=.164\textwidth, trim={0 0cm 1.5cm 0cm},clip]{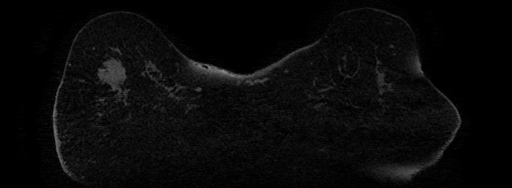}%\hfill
    \includegraphics[height=2.25cm, width=.164\textwidth, trim={0 0cm 1.5cm 0cm},clip]{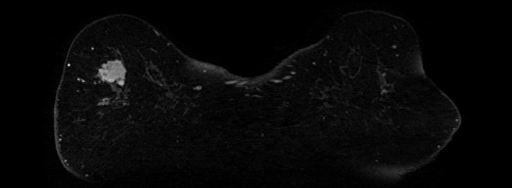}%\hfill
    \includegraphics[height=2.25cm, width=.164\textwidth, trim={0 0cm 1.5cm 0cm},clip]{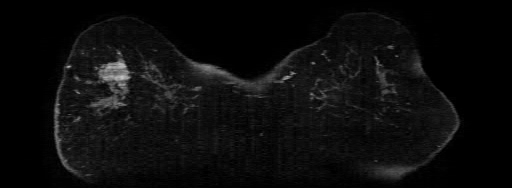}%\hfill
    \includegraphics[height=2.25cm, width=.164\textwidth, trim={0 0cm 1.5cm 0cm},clip]{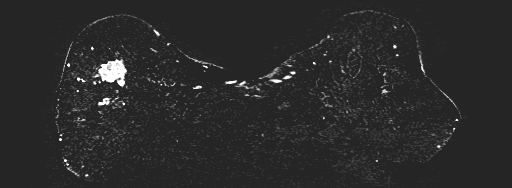}%\hfill
    \includegraphics[height=2.25cm, width=.164\textwidth, trim={0 0cm 1.5cm 0cm},clip]{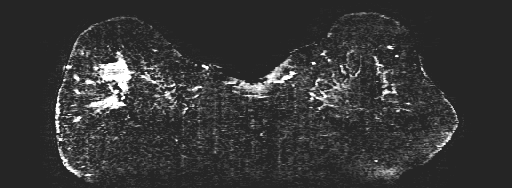}%\hfill
    \includegraphics[height=2.25cm, width=.164\textwidth, trim={0 0cm 1.5cm 0cm},clip]{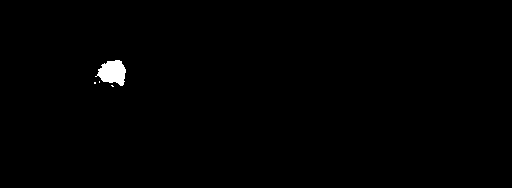}
    \\[\smallskipamount]
    
    \includegraphics[height=2.25cm, width=.164\textwidth, trim={0 4.5cm 0 2.5cm},clip]{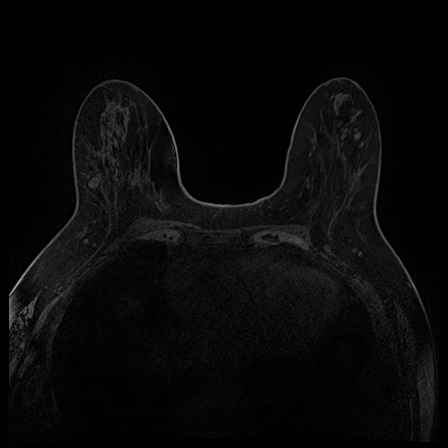}%\hfill
    \includegraphics[height=2.25cm, width=.164\textwidth, trim={0 4.5cm 0 2.5cm},clip]{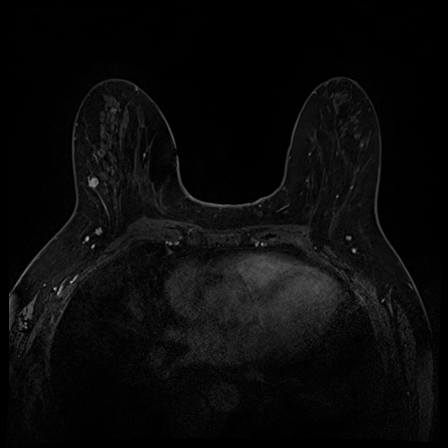}%\hfill
    \includegraphics[height=2.25cm, width=.164\textwidth, trim={0 4.5cm 0 2.5cm},clip]{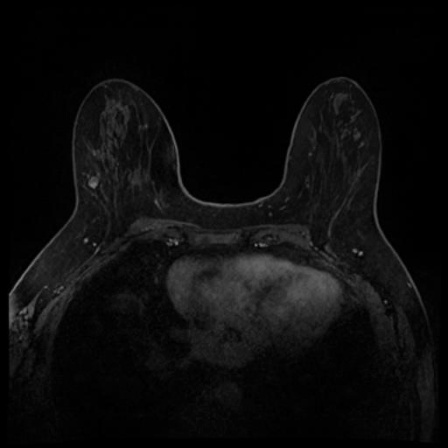}%\hfill
    \includegraphics[height=2.25cm, width=.164\textwidth, trim={0 4.5cm 0 2.5cm},clip]{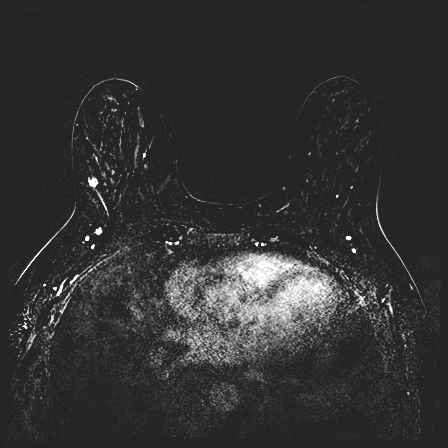}%\hfill
    \includegraphics[height=2.25cm, width=.164\textwidth, trim={0 4.5cm 0 2.5cm},clip]{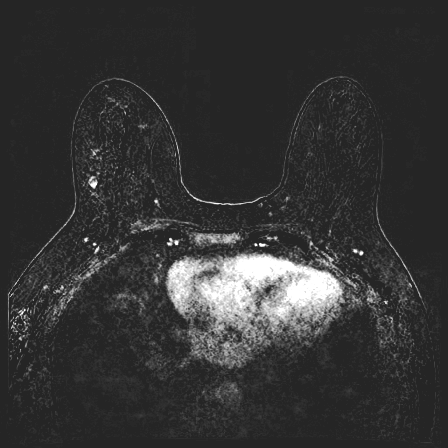}%\hfill
    \includegraphics[height=2.25cm, width=.164\textwidth, trim={0 4.5cm 0 2.5cm},clip]{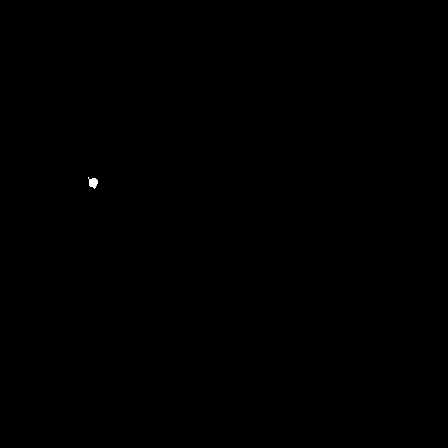}
    %\\[\smallskipamount]
    
    \includegraphics[height=2.25cm, width=.164\textwidth, trim={0 6cm 0 4cm},clip]{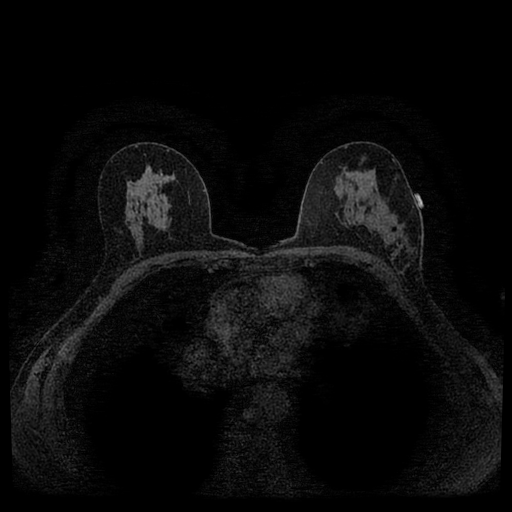}%\hfill
    \includegraphics[height=2.25cm, width=.164\textwidth, trim={0 6cm 0 4cm},clip]{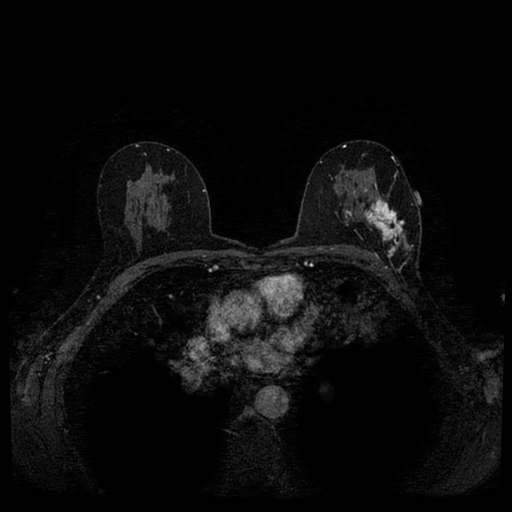}%\hfill
    \includegraphics[height=2.25cm, width=.164\textwidth, trim={0 6cm 0 4cm},clip]{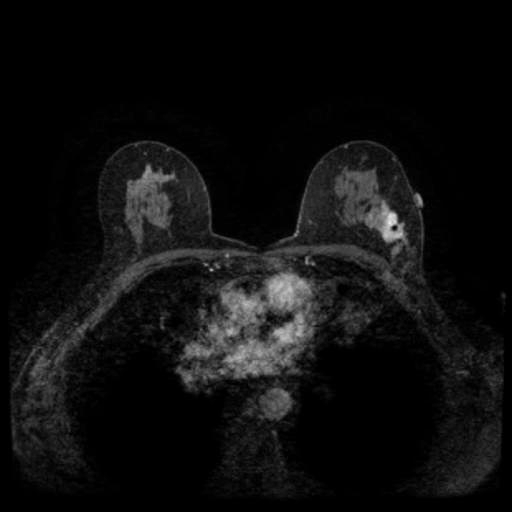}%\hfill
    \includegraphics[height=2.25cm, width=.164\textwidth, trim={0 6cm 0 4cm},clip]{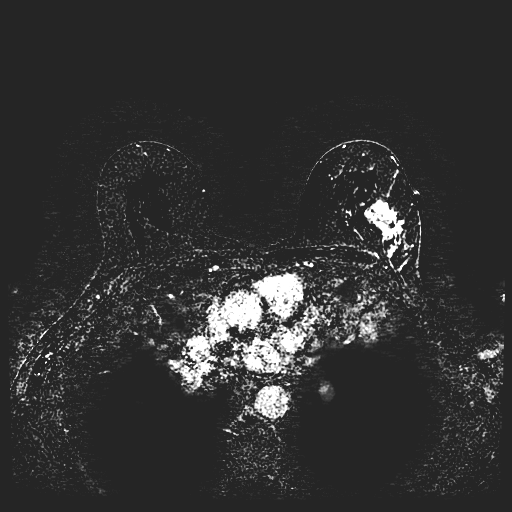}%\hfill
    \includegraphics[height=2.25cm, width=.164\textwidth, trim={0 6cm 0 4cm},clip]{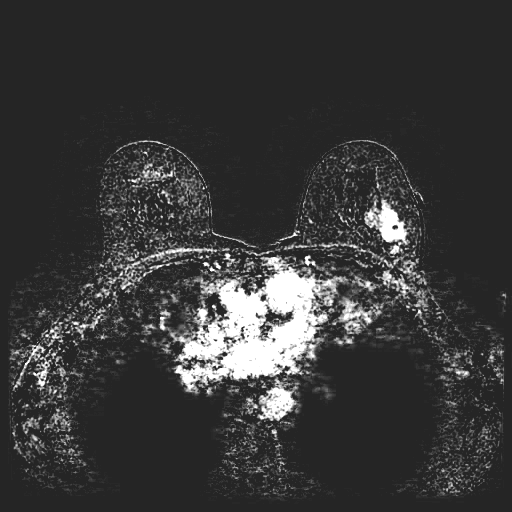}%\hfill
    \includegraphics[height=2.25cm, width=.164\textwidth, trim={0 6cm 0 4cm},clip]{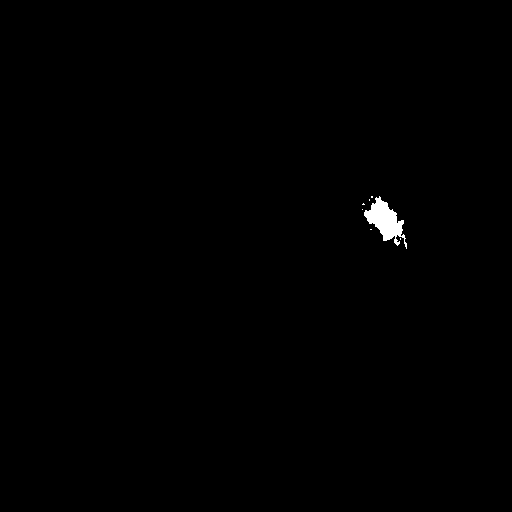}
    %\\[\smallskipamount]
    
    \includegraphics[height=2.25cm, width=.164\textwidth, trim={0 6.5cm 0 1.15cm},clip]{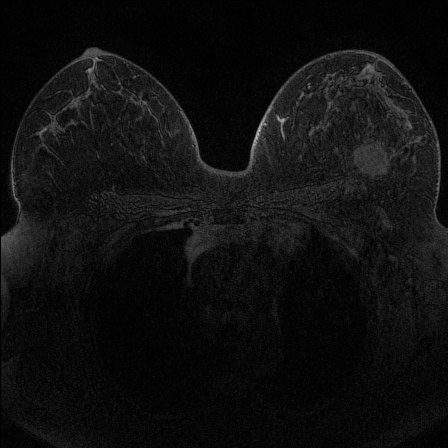}%\hfill
    \includegraphics[height=2.25cm, width=.164\textwidth, trim={0 6.5cm 0 1.15cm},clip]{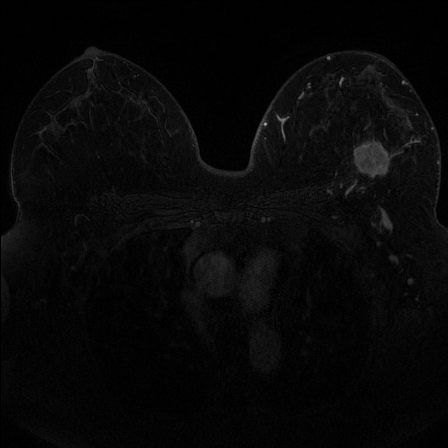}%\hfill
    \includegraphics[height=2.25cm, width=.164\textwidth, trim={0 6.5cm 0 1.15cm},clip]{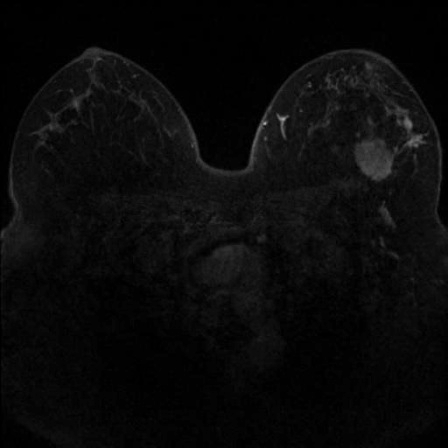}%\hfill
    \includegraphics[height=2.25cm, width=.164\textwidth, trim={0 6.5cm 0 1.15cm},clip]{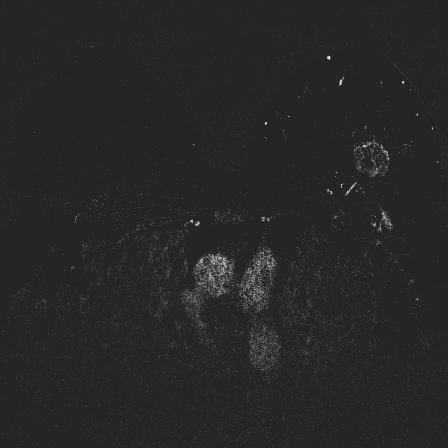}%\hfill
    \includegraphics[height=2.25cm, width=.164\textwidth, trim={0 6.5cm 0 1.15cm},clip]{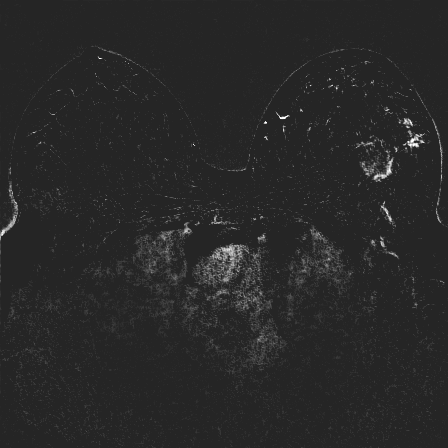}%\hfill
    \includegraphics[height=2.25cm, width=.164\textwidth, trim={0 6.5cm 0 1.15cm},clip]{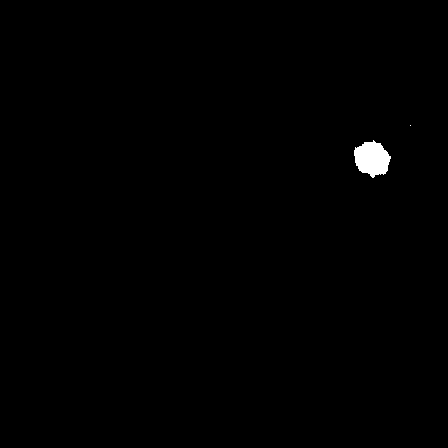}

    \includegraphics[height=2.25cm, width=.164\textwidth, trim={1.5cm 6.5cm 1.5cm 3cm},clip]
    {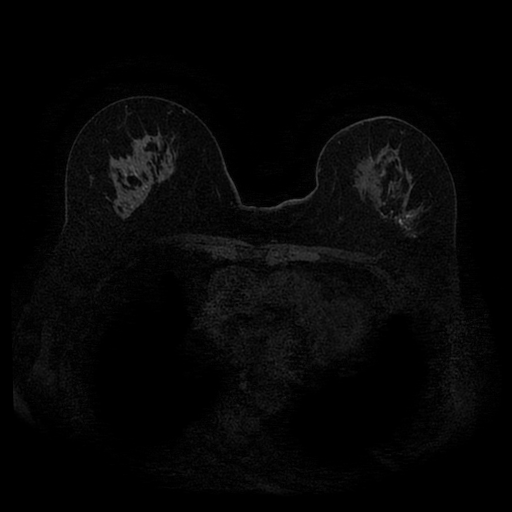}%\hfill
    \includegraphics[height=2.25cm, width=.164\textwidth, trim={1.5cm 6.5cm 1.5cm 3cm},clip]
    {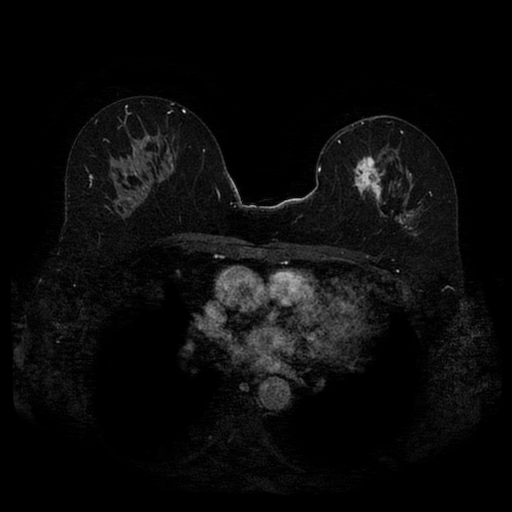}%\hfill
    \includegraphics[height=2.25cm, width=.164\textwidth, trim={1.5cm 6.5cm 1.5cm 3cm},clip]
    {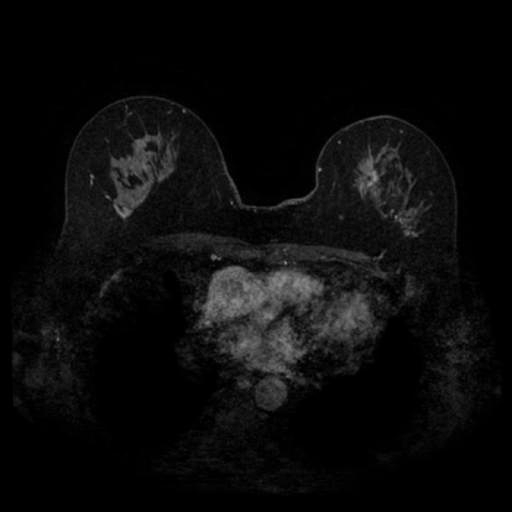}%\hfill
    \includegraphics[height=2.25cm, width=.164\textwidth, trim={1.5cm 6.5cm 1.5cm 3cm},clip]{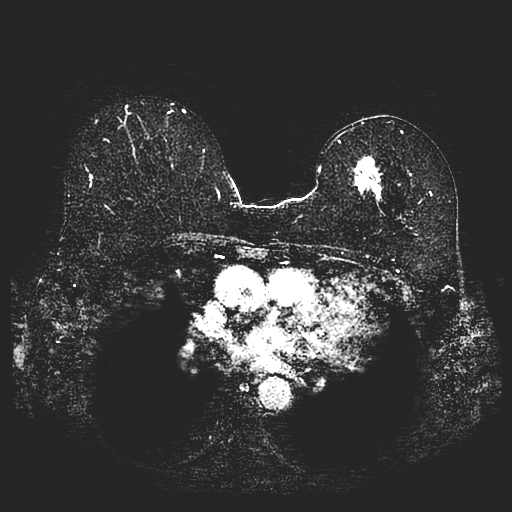}%\hfill
    \includegraphics[height=2.25cm, width=.164\textwidth, trim={1.5cm 6.5cm 1.5cm 3cm},clip]{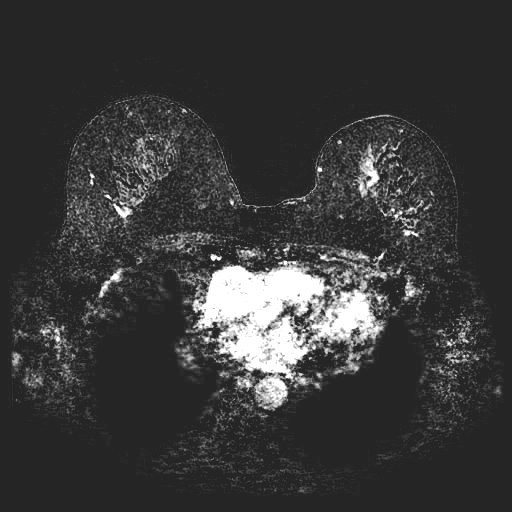}%\hfill
    \includegraphics[height=2.25cm, width=.164\textwidth, trim={1.5cm 6.5cm 1.5cm 3cm},clip]
    {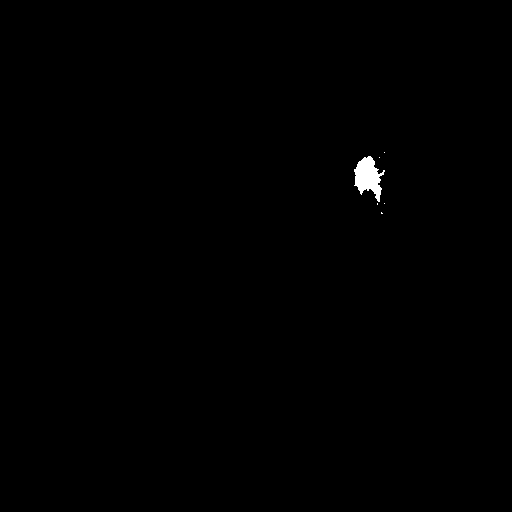}

    \includegraphics[height=2.25cm, width=.164\textwidth, trim={1.5cm 6.5cm 1.5cm 3.2cm},clip]{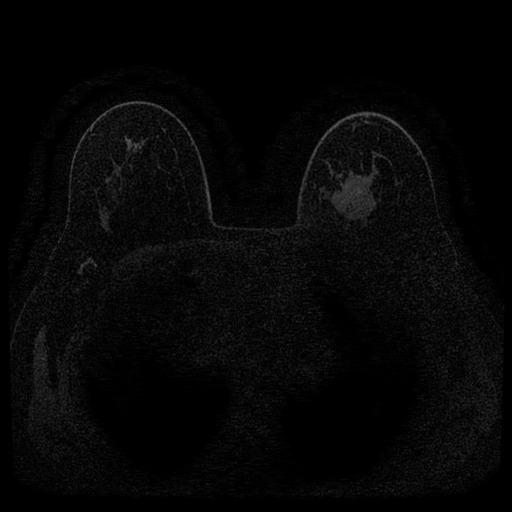}%\hfill
    \includegraphics[height=2.25cm, width=.164\textwidth, trim={1.5cm 6.5cm 1.5cm 3.2cm},clip]{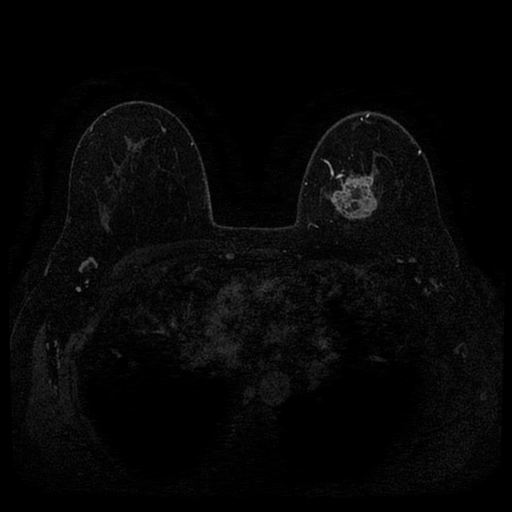}%\hfill
    \includegraphics[height=2.25cm, width=.164\textwidth, trim={1.5cm 6.5cm 1.5cm 3.2cm},clip]
    {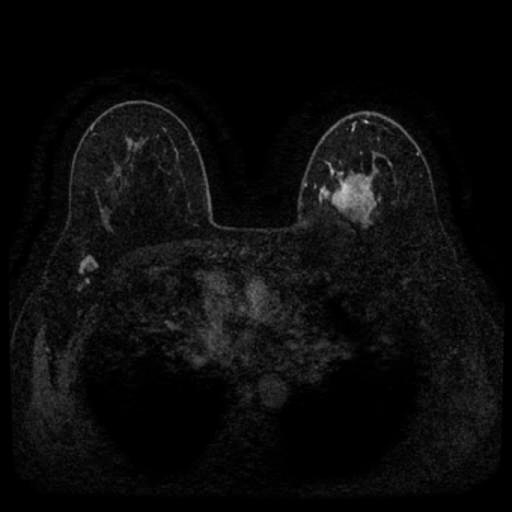}%\hfill
    \includegraphics[height=2.25cm, width=.164\textwidth, trim={1.5cm 6.5cm 1.5cm 3.2cm},clip]{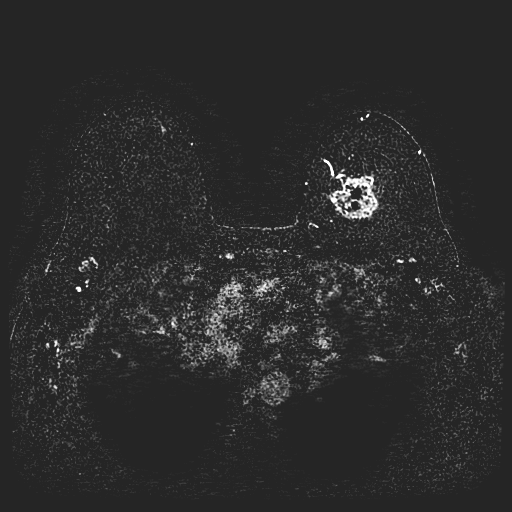}%\hfill
    \includegraphics[height=2.25cm, width=.164\textwidth, trim={1.5cm 6.5cm 1.5cm 3.2cm},clip]{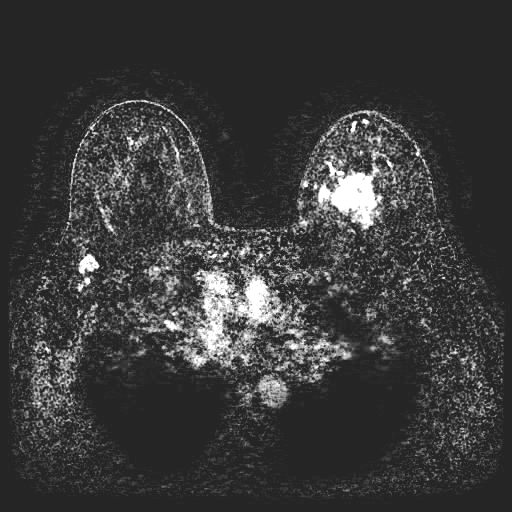}%\hfill
    \includegraphics[height=2.25cm, width=.164\textwidth, trim={1.5cm 6.5cm 1.5cm 3.2cm},clip]
    {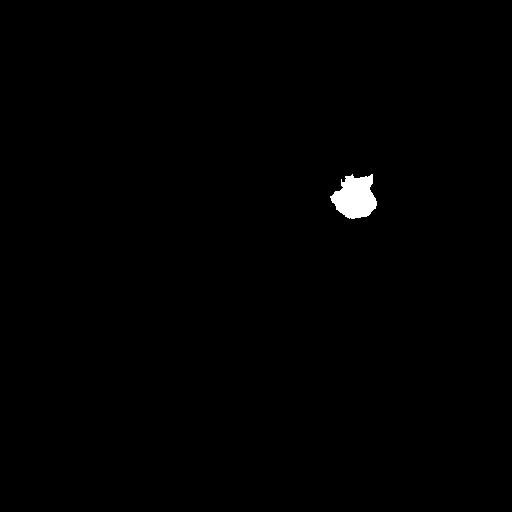}
    \caption{Breast DCE-MRI synthesis, as shown for six cases from the Duke Dataset \cite{saha2018machine}, two of which are manually selected from the validation set (\nth{1}-\nth{3} row: Case 228, \nth{4} row: Case 886), two manually selected from the test set (\nth{5} row: Case 378, \nth{6} row: Case 907), and two randomly selected from the test set (\nth{7} row: Case 041, \nth{8} row: Case 045). From the left to the right, the corresponding axial slices are depicted for (\emph{a}) the real pre-contrast, (\emph{b}) the real post-contrast phase 1, (\emph{c}) the synthetic post-contrast phase 1, (\emph{d}) the subtraction image based on the real post-contrast image, (\emph{e}) the subtraction image based on the synthethic post-contrast image, and (\emph{f}) the ground truth segmentation mask. Intensity and contrast of the subtraction images was increased using OpenCV\cite{opencv} (same scaling for all images). Samples of case 228 are shown in the axial (\nth{1} row, slice 111), saggital (\nth{2} row, slice 119), and coronal view (\nth{3} row, slice 286). The synthetic images from the coronal and sagittal planes are extracted from the respective 3D volume that is based on stacked synthetic 2D axial slices.}
 \label{fig:image_comparison}
 \end{figure}
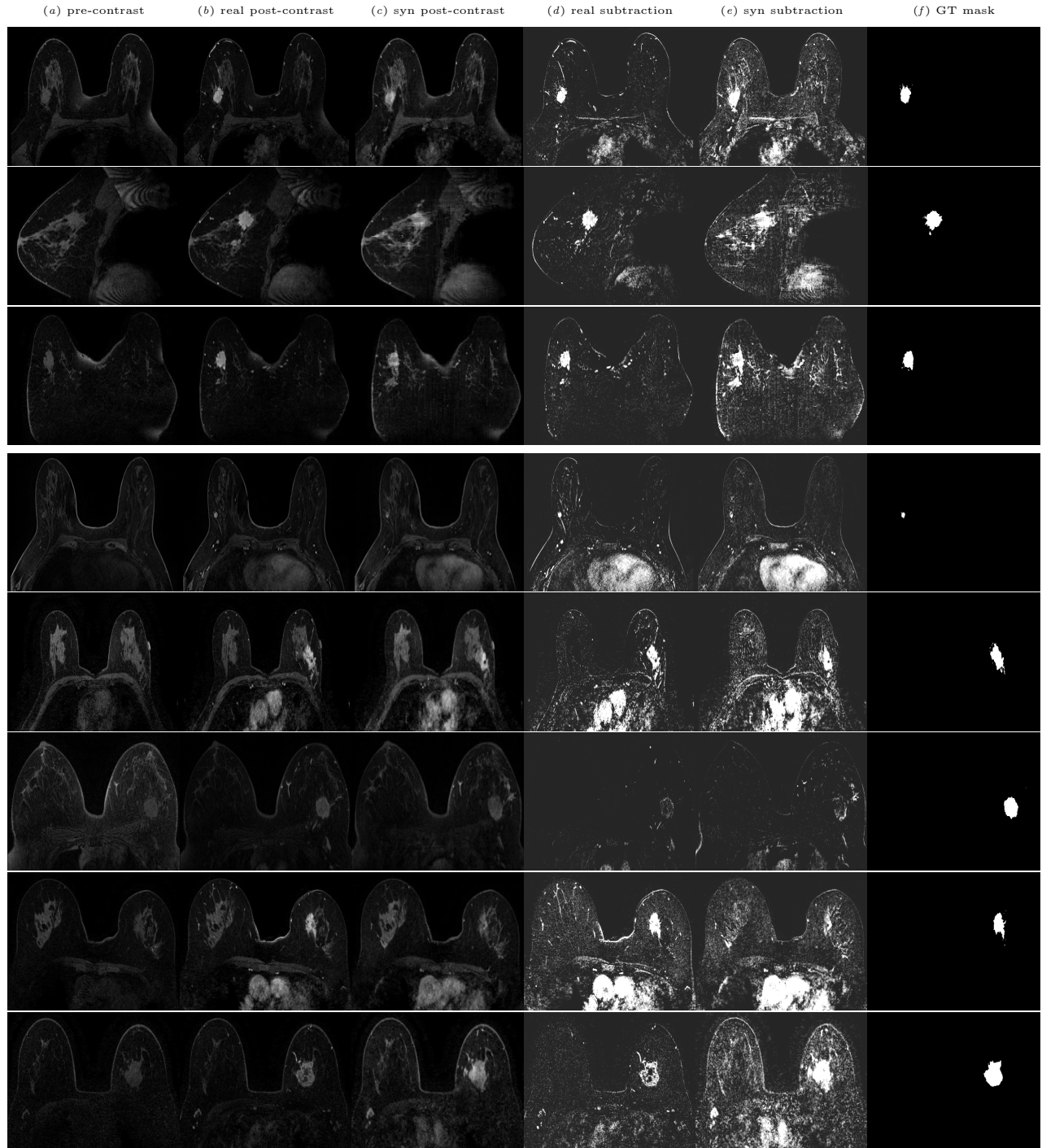

To segment tumours as 3D volumes, we adopt a single 3D U-Net\cite{ronneberger2015u} model   based on the nnU-Net framework \cite{isensee2021nnu} (\textit{nnunetv2 3d full\_res}), however, without applying any of nnU-Net's post-processing techniques.
The segmentation model is trained for 500 epochs for each fold in a 5-fold cross-validation (CV). The performance on the test set is obtained from the average predictions of the ensemble of the five trained CV models.
As the ground truth segmentations only contain the primary lesion, we opted to remove multifocal cases (33 cases) from the dataset. Furthermore, we crop the images to encompass only a single breast per image, thus avoiding any bilateral cases. Bias field correction \cite{tustison2010n4itk} is applied, and the same segmentation masks are used for real and synthetic data, as illustrated in  Figure \ref{fig:segmentation} and Figure \ref{fig:image_comparison}.
Segmentation is evaluated using the Dice coefficient, which 
%is depicted in Equation \ref{eq:3} and 
%quantifies the overlap between the predicted segmentation %(A) 
%and the ground truth. %(B). 
%The Dice score 
ranges between [0, 1], with 0 representing no overlap, while 1 indicates a complete overlap between predicted volume and ground truth tumour volume.
%\begin{equation} \label{eq:3}
%    Dice = 2\frac{\left \| A \cap B  \right \|}{\left \| A \right \| + \left \| B \right \|}
%\end{equation}

\subsection{Metrics and Model Selection}
%\subsubsection{Image Distribution Similarity Metrics}
\paragraph{Image Quality Metrics} To perceptually compare the distributions of two image datasets, we adopt the Fréchet Inception Distance (FID) \cite{heusel2017gans}. The FID is based on the distance between features of two datasets extracted via a pretrained Inception \cite{szegedy2016rethinking} model.  Following \cite{osuala2023medigan}, we adopt both the standard Inception v1 feature extractor pretrained on ImageNet\cite{deng2009imagenet} and a radiology domain-specific inception v3 feature extractor pretrained on the RadImageNet\cite{mei2022radimagenet} dataset, as FID$_{Img}$ and FID$_{Rad}$, respectively. 
The latent features are extracted from both synthetic and real datasets before being fitted to multi-variate Gaussians $X$=real and $Y$=synthetic with means $\mu_{X}$ and $\mu_{Y}$ and covariance matrices $\Sigma_{X}$ and $\Sigma_{Y}$. Lastly, the distance between $X$ and $Y$ is computed as the Fréchet distance depicted in Equation \ref{eq:2}.
    \begin{equation} \label{eq:2}
    \begin{aligned}
        %\textit{FID}(X, Y, M) = 
        \textit{FD}(X, Y) = \lVert \mu_{X} - \mu_{Y} \rVert_{2}^{2} + \text{tr}(\Sigma_{X} + \Sigma_{Y} -2(\Sigma_{X}\Sigma_{Y})^{\frac{1}{2}})
    \end{aligned}
    \end{equation}
%\subsubsection{Image Similarity Metrics}
For pixel-level image pair similarity comparison, we adopt the mean squared error (MSE) and the mean absolute error (MAE). These metrics compute the (squared) difference per corresponding pixels between a real and synthetic image pair. This error is averaged across all pixels of an imagepair and, lastly, averaged across all image pairs.
Furthermore, we adopt the structural similarity index measure (SSIM) \cite{wang2004image} as a further popular \cite{osuala2022data} medical synthetic data evaluation metric on image pair level. 
SSIM\cite{wang2004image} measures the perceived image quality based on a combination of luminance, contrast, and structural information. As before, we average the SSIM across real and synthetic sample-pairs. 

\paragraph{Scaled Aggregate Measure (SAMe)} In our analysis of the related literature \cite{osuala2022data, muller2023using, wang2021synthesizing,xue2022bi,kim2022tumor, zhang2023synthesis,zhao2020tripartite, pasquini2022synthetic}, 
we observe that there is no consensus on which metrics to use when evaluating the quality of synthetic data in image-to-image synthesis tasks. While different metrics are used and reported, it is unclear which metric to prioritise when metrics provide contrary information. Beyond quality assessment, this problem extends to the question of which metric to use to assess when to stop training a generative model.
%, where the same metrics are used to asses which generative model training checkpoint to select for inference application.
As different aspects of truth are present in each metric, we propose an ensemble of metrics to be the best approach 
to evaluate synthetic data. 
To this end, we introduce the Scaled Aggregate Measure (SAMe) that scales several metrics, in this work, namely, the SSIM, MSE, MAE, FID$_{Img}$\cite{deng2009imagenet,heusel2017gans}, and FID$_{Rad}$\cite{mei2022radimagenet,osuala2023medigan}
to $[0, 1]$ using per-metric min-max normalisation, where smaller values
indicate better performance (SSIM was reversed after scaling). SAMe is then calculated as the average between these metrics. The choice of the metrics in SAMe is based on a balanced selection of perceptual metrics that capture the global semantics (FID) and perceived quality of images (SSIM and FID) and on fine-grained pixel-level comparison metrics (MAE and MSE) that assess the accurateness of replication between an image pair. While FID has a high sensitivity to small changes and close relevance to human inspection~\cite{borji2022}, SAMe's pixel space metrics measure objective (MSE, MAE) and subjective image fidelity (SSIM)~\cite{samajdar2015}.
SAMe combines analytical measures (SSIM, MAE, MSE) with latent features of neural networks (FID), with the latter being further divided into domain-agnostic (FID$_{Img}$) and radiology domain-specific (FID$_{Rad}$) features to capture different pieces of relevant information in the evaluated synthetic images. This allows SAMe to combine complementary and mutually exclusive information present in the selected image quality metrics into a single meaningful measure.

\section{RESULTS}
\label{sec:results}

\subsection{Synthetic Data Quality Assessment}

\begin{figure} [ht]
\begin{center}
\includegraphics[height=5cm]{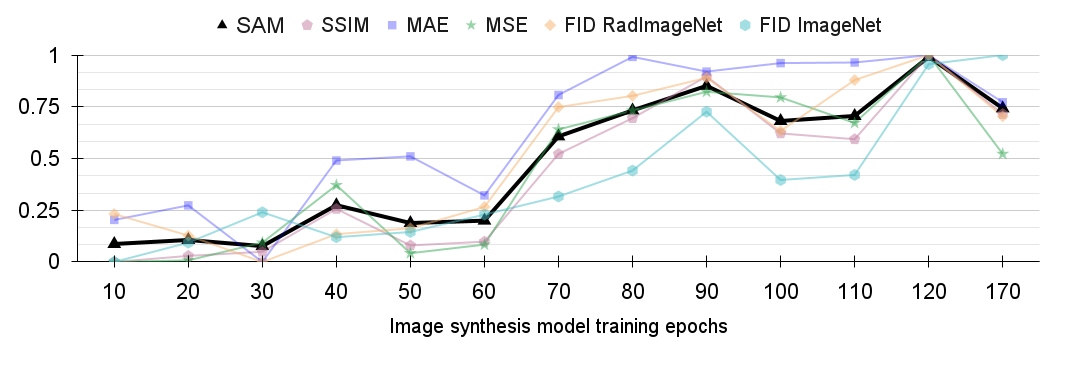}
\end{center}
\vspace{-6mm}
\caption[metrics] 
{\label{fig:metrics} Synthetic image distribution (FID$_{Img}$ and FID$_{Rad}$), and pixel space objective (MSE and MAE) and perception-based (SSIM) quality metrics across the generative models' training epochs. Utilising these, we introduce the Scaled Aggregate Measure (SAMe) to inspect the overall quality (the lower, the better) and enabling an informed selection of the best training checkpoint (i.e., epoch 30, achieving the lowest SAMe)
for image generation.
FID metrics are computed for 3000 and MSE, MAE and SSIM metrics for 5000 synthetic-real post-contrast pairs of axial MRI slices from the validation set.
% Image quality metrics on the validation set.
}
\end{figure}

\begin{figure}
  \centering
  \includegraphics[width=\textwidth]{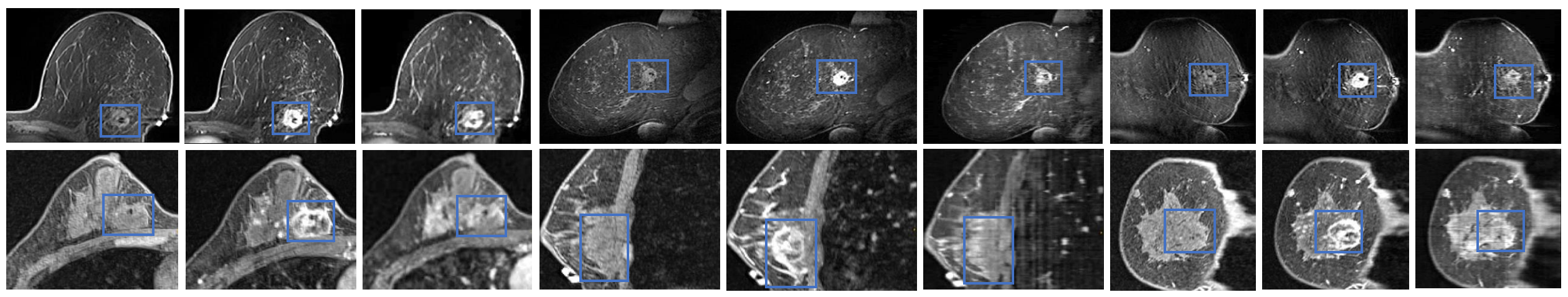}
\caption{Single breast examples of cropped T1 MRI slices with tumour bounding box for two randomly selected test cases from the Duke Dataset \cite{saha2018machine}. Case 612 (normal) is shown in the top row and Case 005 (difficult) in the bottom row. Each row is organised in 3 by 3 columns (order: axial, sagittal, coronal), where the first, second, and third column corresponds to 
pre-contrast, real post-contrast, and synthetic post-contrast, respectively. The intensity of these images was auto-adjusted using ITKSnap \cite{yushkevich2016itk}.
}
\label{fig:cropped_images}
\end{figure}

After training the generative model, we generate T1-weighted DCE-MRI phase 1 images (often corresponding to peak enhancement in the studied dataset\cite{saha2018machine}) for the image synthesis test (30 cases) and validation (224 cases) sets. %using paired image-to-image translation, where the input was a pre-contrast MRI slice along with its real post-contrast counterpart at the selected phase, and the output a synthetic post-contrast image. 
Qualitative results of our image synthesis model, based on its translation of entire axial breast MRI slices to the post-contrast domain, are depicted in Figures \ref{fig:image_comparison} and \ref{fig:cropped_images}.
Figure \ref{fig:image_comparison} visualises qualitative pre- to post-contrast translation results alongside respective subtraction images for six different patient cases. Figure \ref{fig:cropped_images} focuses on cropped views of the region-of-interest showing the tumour area in the axial, sagittal and coronal plane of two randomly selected patient cases. The top row illustrates a case of average difficulty while the bottom row shows a particularly difficult pre- to post-contrast translation case. We note that some hallucinations of false-positive contrast regions exist (e.g., see difficult case in the bottom row of Figure \ref{fig:cropped_images}) and that some tumours are only partly contrast-enhanced (e.g., see synthetic post-contrast image of randomly selected case 041 in the \nth{7} row of Figure \ref{fig:image_comparison}). In randomly selected case 045 shown in the \nth{8} row of Figure \ref{fig:image_comparison}, the real post-contrast image illustrates hypointense regions within the tumour indicating a necrotic tumour core. With limited observability of this manifestation in the pre-contrast domain, this characteristic is not translated into the synthetic post-contrast domain despite successful tumour localisation demonstrated by the synthetic subtraction image. Overall, our model's qualitative results demonstrate encouraging pre- to post-contrast image translation coupled with promising synthetic contrast localisation and injection capabilities. 

To further systematically evaluate the image synthesis quality, we examine quality comparison metrics between synthetic and real post-contrast MRI slices. 
% These included fr\'echet inception distance (FID), mean squared error (MSE), mean absolute error (MAE), and structural similarity index measure (SSIM) \comment{on a slice level, right?}. For FID, we calculated the distances using pre-trained Inception v3 and v1 weights on ImageNet (FID-ImageNet) and RadImageNet (FID-RadImageNet) respectively for the same 3000 slices. These slices correspond to \dots \comment{how many cases, or were randomly selected, or? }.  
%These cover the generated distribution and image samples, while 
%To this end, we apply SAMe to guide the generative model checkpoint selection. %Specifically, FID accounts for measuring the generated distribution's distance from the target real one, with a high sensitivity to small changes and close relevance to human inspection~\cite{borji2022}. %The rest of the metrics act on the pixel space of the generated samples, measuring objective (MSE, MAE) and subjective fidelity (SSIM)~\cite{samajdar2015}. 
%All metrics were calculated on a validation set of 224 cases which is separate from the generative training process, as described in section~\ref{sec:methodology}.
%
%
%The metrics are scaled via min-max normalisation to a range of $[0, 1]$ for demonstration and comparison purposes. Smaller values indicate better performance (SSIM was reversed after scaling). 
As shown in Figure ~\ref{fig:metrics}, across and within training epochs, the different metrics provide different inconsistent conclusions regarding the best performing synthesis model - a fact that calls for a unified measure, i.e. SAMe. On the other hand, all metrics follow a similar general trend, maintaining lower (better) values until epoch 60, after which they start to rise substantially, indicating a potential overfitting and no further benefit of additional training. By choosing the model with the lowest SAMe, we are able to determine epoch 30 as the optimal model checkpoint to generate the synthetic post-contrast samples that are further used in the tumour segmentation downstream task. Table \ref{tab:image_quality} summarises computed synthetic data quality metrics alongside their standard deviation on validation and test set.
\begin{table}[ht!]
\centering
\caption{Synthetic image quality evaluation based on metrics SAMe, FID$_{Img}$, FID$_{Rad}$, SSIM, MAE, MSE. The reported results (with standard deviation, where applicable) are based on 3000 (FID) or 5000 (MSE, MAE and SSIM) synthetic-real post-contrast image pairs of axial MRI slices. The synthetic images, where e.g. \emph{Syn$_{ep30}$} images are generated after 30 epochs of GAN training, are compared to their corresponding real post-contrast counterparts. \emph{Real Post vs. Real Pre} indicates the comparison between corresponding real pre- and post-contrast images. \emph{Subt} refers to subtraction images, where  pre-contrast images are subtracted from either their real (\emph{Real Subt}) or synthetic (\emph{Syn Subt}) post-contrast counterpart. \emph{Splitted Test} refers to a random by-patient split of the test set (i.e. without corresponding image pairs) capturing the variance across patient cases. As the test set contains 5186 images, the number of image pairs for FID computation in test was reduced to 2000.}
% \scriptsize
\vspace{2mm}
\resizebox{1.0\columnwidth}{!}{
\begin{tabular}{lccccccc}
    \toprule
     & & \multicolumn{6}{c}{Metric} \\
     \arrayrulecolor{light_grey} \cmidrule(lr){3-8}
    Comparison & Dataset &  FID$_{Img}$ $\downarrow$ & FID$_{Rad}$ $\downarrow$ & SSIM $\uparrow$ & MAE $\downarrow$ & MSE $\downarrow$ & SAMe $\downarrow$ \\
    \arrayrulecolor{black} \toprule
    Real Post vs. Syn$_{ep10}$ Post & Val & \textbf{15.047} & 0.108 & \textbf{0.701(0.081)} & 93.895(41.748) & \textbf{37.803(9.960)} & 0.087 \\
    Real Post vs. Syn$_{ep30}$ Post    & Val & 17.308 & \textbf{0.081} & 0.699(0.081) & 88.733(39.426) & 38.334(9.582) & \textbf{0.077} \\
    Real Post vs. Syn$_{ep50}$ Post    & Val & 16.412 & 0.089 & 0.696(0.090) & 101.696(44.672) & 38.045(10.985) & 0.188 \\
    Real Post vs. Syn$_{ep100}$ Post   & Val & 18.778 & 0.219 & 0.669(0.116) & 113.144(59.360) & 42.320(17.792) & 0.682 \\
    Real Post vs. Real Pre & Val & 34.062 & 0.120 & 0.660(0.090) & \textbf{66.146(31.758)} & 42.933(11.528)& \\
    \arrayrulecolor{light_grey} \cmidrule(lr){1-8}
    Real Post vs. Syn$_{ep30}$ Post   & Test &  \textbf{28.717} & \textbf{0.0385} & \textbf{0.726(0.089)} & 85.623(38.297) & \textbf{34.882(10.520)} \\
    Real Post vs. Real Pre                 & Test & 59.644 & 0.1556  & 0.705(0.104) & \textbf{66.121(34.473)} & 40.124(16.183) \\
    \arrayrulecolor{light_grey} \cmidrule(lr){1-8}
    Real Subt vs. Syn$_{ep30}$ Subt  & Test & 46.931 & 0.2864 & 0.692(0.097) & 44.896(23.403) & 23.425(8.602) \\
    \arrayrulecolor{light_grey} \cmidrule(lr){1-8}
    Real Post vs. Syn$_{ep30}$ Post & Splitted Test & \textbf{43.865} & 0.7012 & & & &\\
    Real Post vs. Real Post              & Splitted Test  & 49.808  & \textbf{0.2060} & & & &\\
    \arrayrulecolor{black} \bottomrule
\end{tabular}
}
\label{tab:image_quality}
% \vspace{-2mm}
\end{table}
Table \ref{tab:image_quality} further compares the 2D full axial slice image dataset similarity between the same synthetic and real test case images. In this comparison, the synthetic post-contrast images are semantically (FID scores) and perceptually (SSIM) substantially closer to the real post-contrast images than the real pre-contrast images. 
We note that in the comparison of \emph{splitted test} datasets, the compared sets do not correspond to the same patient cases. This allows to measure the variability across patient test cases. Interestingly, based on the domain-agnostic FID$_{Img}$ the variability between real post-contrast cases results higher than the one between real and synthetic post-contrast cases. On the contrary, the FID$_{Rad}$ shows, for the same dataset split, less variability between real post-contrast datasets than between real and synthetic ones. Specifically according to the radiology domain-specific FID$_{Rad}$, the variability across patient cases (\emph{splitted test}) is estimated as being generally higher than the variability between pre-, post- and synthetic post-contrast sequences (\emph{test}) of corresponding cases. We further analyse subtraction images, which are created by subtracting a pre-contrast image from either its real or synthetic post-contrast counterpart. In this regard, comparing corresponding real (\emph{Real Subt}) with synthetic (\emph{Syn$_{ep30}$ Subt}) subtraction images results in improved reconstruction-based metrics (i.e. MSE, MAE) compared to the comparison of real vs. synthetic post-contrast images. However, this improvement can be attributed to the clipping of pixel values to 0 when their value resulted negative after subtraction. In terms of perceptual (e.g., SSIM) and latent feature distribution-based metrics (e.g., FID$_{Rad}$, FID$_{Img}$) the real vs. synthetic post-contrast image comparison achieves better quantitative results than its subtraction image equivalent.

For reference and comparison, we extend on Table \ref{tab:image_quality} computing several additional metrics, namely the peak signal to noise ratio (PSNR), the multiscale structural similarity (MS-SSIM) \cite{wang2003multiscale} and the Learned Perceptual Image Patch Similarity (LPIPS) \cite{zhang2018unreasonable}. These metrics are computed on 5000 test set image pairs for (a) real vs. syn$_{ep30}$ post-contrast, for (b) real post-contrast vs. real pre-contrast, and (c) real vs. syn$_{ep30}$ subtraction image pairs. The metric values alongside their standard deviation for (a) are 
PSNR$\uparrow$=32.91(1.35), MS-SSIM$\uparrow$=0.798(0.08), 
LPIPS$\downarrow$=0.064(0.04), for
(b) they are 
PSNR$\uparrow$=32.42(1.68), 
MS-SSIM$\uparrow$=0.780(0.07), LPIPS$\downarrow$=0.084(0.05), and for (c) they resulted in 
PSNR$\uparrow$=34.74(1.73), 
MS-SSIM$\uparrow$=0.717(0.09), LPIPS$\downarrow$=0.062(0.03).

\subsection{Tumour Segmentation Experiments}
Under the assumption of different data access limitations in the pre- and post-contrast domain, we perform four blocks of tumour segmentation experiments. 
The \nth{1} block in Table~\ref{tab:experiments_1_2} assumes post-contrast data is not available for segmentation training or testing. Available pre-contrast training cases (\emph{baseline 1}) are augmented with their synthetic post-contrast equivalents.
The \nth{2} set of experiments in Table~\ref{tab:experiments_1_2} assumes pre-contrast data is available only for training, while the test data distribution consists of post-contrast images. Assessing tumour segmentation performance under this domain shift, we evaluate the effect of adding synthetic post-contrast cases to pre-contrast \emph{baseline 2}. 
The \nth{3} block in Table~\ref{tab:experiments_3_4} describes the clinical scenario where real post-contrast data is available and used for both training and testing assessing whether synthetic data can improve upon post-contrast \emph{Baseline 3}.
The \nth{4} set of experiments in Table~\ref{tab:experiments_3_4} analyses the scenario where segmentation models are trained on real post-contrast data, but tested on pre-contrast cases. This includes the cases where contrast agents are not administered e.g., in specific patient sub-populations such as pregnant or kidney-compromised patients, patients who refuse contrast media injection, or patients with significant risks of allergic reactions to contrast media.
\begin{table}[ht!]
\centering
\caption{Results for tumour segmentation test cases. Experiments \emph{1} show the scenario without contrast agent administration, where segmentation training and testing data is within the pre-contrast domain. 
%The synthetic data is either added as training data augmentation, or as second channel input into the segmentation model alongside the corresponding real pre-contrast data, or as synthetic post-contrast minus real pre-contrast subtraction image. 
Experiments \emph{2} analyse the case where a model has been trained without (access to) real contrast-enhanced data, but is nevertheless applied on test cases in the post-contrast domain (domain shift). %The reported Dice coefficient results are derived from the ensemble model over 5-fold cross-validation.
}
% \scriptsize
\vspace{2mm}
\resizebox{0.95\columnwidth}{!}{
\begin{tabular}{lc}
    \toprule
     Experiments \emph{1}. \small{Pre-contrast training data and pre-contrast test data available} &  Dice $\uparrow$ \\ \emph{train on:} & \emph{test on:} Real pre-contrast  \\
    %\arrayrulecolor{black} \toprule
    \arrayrulecolor{light_grey} \toprule
    Real pre-contrast (\emph{baseline 1}) & \textbf{0.569}  \\
    Real pre-contrast + syn post-contrast (augmentation) & 0.531  \\
    %Real pre-contrast + syn post-contrast (2-channels) & tbd \\
    %Real pre-contrast + syn post-contrast (subtraction) & tbd \\
    % 
    Syn post-contrast & 0.486 \\
    %Real post-contrast (\emph{lower bound}) & 0.164 \\
    %\arrayrulecolor{black} \bottomrule
    %\arrayrulecolor{black} \toprule
    \arrayrulecolor{black} \toprule
     Experiments \emph{2}. \small{Domain shift: Pre-contrast training data, but no pre-contrast test data available}  &  \\ \emph{train on:} & \emph{test on:} Real post-contrast  \\
    \arrayrulecolor{light_grey} \toprule
    Real pre-contrast (\emph{baseline 2}) & 0.484  \\
    Real pre-contrast + syn post-contrast (augmentation) & 0.663  \\
    %Real pre-contrast + syn post-contrast (2-channels) & tbd \\
    %Real pre-contrast + syn post-contrast (subtraction) & tbd \\
    % 
    Syn post-contrast & \textbf{0.687} \\
    \arrayrulecolor{black} \bottomrule
\end{tabular}
}
\label{tab:experiments_1_2}
% \vspace{-2mm}
\end{table}
In all data augmentation experiments, for each case in the training data, the respective augmented version of that case (e.g., the real and/or synthetic post-contrast volume) is added to the training data. We note that the model is not provided with information that an initial training data point (e.g., a pre-contrast volume) and its augmented equivalent (e.g., a synthetic post-contrast volume) correspond to the same patient case.
The reported Dice coefficients are derived as ensemble predictions of the five segmentation models trained in the 5-fold cross validation\cite{isensee2021nnu}. For this reason, no standard deviation is reported.

Inspecting \emph{baseline 1} in Table \ref{tab:experiments_1_2}, we note that 
synthetic post-contrast augmentations do not improve segmentation performance in the pre-contrast domain. However, in the domain shift scenario of \emph{baseline 2}, the synthetic data augmentations do result in a substantial improvement in the post-contrast domain. Specifically, training with real pre-contrast augmented by synthetic post-contrast images improves the post-contrast Dice coefficient by 
%$36.98\%$ 
$0.179$ (i.e., from $0.484$ to $0.663$)
compared to the baseline while maintaining a comparative level of performance in the pre-contrast domain (i.e., $0.531$ as compared to $0.569$). 
%This is $32.85\%$ more than the real post-contrast augmentations, while real and synthetic post-contrast augmentations achieve an improvement of \comment{TBU}$X\%$.
This result corroborates the image quality analysis findings in Table \ref{tab:image_quality} that show that the synthesised images are within the post-contrast domain distribution and further shows its usefulness in domain-shift scenarios.
\begin{table}[ht!]
\centering
\caption{Results for tumour segmentation test cases. In experiments \emph{3} data from both pre-contrast as well as post-contrast acquisitions is available enabling testing in the post-contrast domain.
Experiments \emph{4} demonstrate the benefit of synthetic post-contrast data in the scenario where segmentation models were trained in the post-contrast domain, but need to be applied to test subjects for which only pre-contrast data is available (e.g., due to allergy, pregnancy, or missing consent). %The reported Dice coefficient results are derived from the ensemble model over 5-fold cross-validation.
}
% \scriptsize
\vspace{2mm}
\resizebox{0.95\columnwidth}{!}{
\begin{tabular}{lc}
    \toprule
    Experiments \emph{3}. \small{Post-contrast training data and post-contrast test data available}  & Dice $\uparrow$ \\  \emph{train on:} & \emph{test on:} Real post-contrast  \\
    %\arrayrulecolor{black} \toprule
    \arrayrulecolor{light_grey} \toprule
    Real post-contrast (\emph{baseline 3}) & 0.790 \\
    Real post-contrast + syn post-contrast (augmentation)  & \textbf{0.797} \\
    Real post-contrast + real pre-contrast (augmentation) &  0.780 \\
    Real post-contrast + real pre-contrast + syn post-contrast (augmentation) & 0.770 \\
    Syn post-contrast & 0.687 \\
    % 
    %\arrayrulecolor{black} \bottomrule
    \arrayrulecolor{black} \toprule
     Experiments \emph{4}. \small{Domain shift: Post-contrast training data, but no post-contrast test data available}  & \\ \emph{train on:} &  \emph{test on:} Real pre-contrast   \\
    \arrayrulecolor{light_grey} \toprule
    Real post-contrast (\emph{baseline 4}) & 0.164 \\
    Real post-contrast + syn post-contrast (augmentation)  & 0.409 \\
    Syn post-contrast & \textbf{0.486} \\
    \arrayrulecolor{black} \bottomrule
\end{tabular}
}
\label{tab:experiments_3_4}
% \vspace{-2mm}
\end{table}
\emph{Baseline 3} in Table \ref{tab:experiments_3_4} provides a strong tumour segmentation performance in the post-contrast domain ($0.790$). Nevertheless, synthetic post-contrast augmentation achieves a slight improvement over this baseline ($0.797$) and is preferable to pre-contrast augmentations ($0.780$).
As to \emph{baseline 4}, synthetic post-contrast augmentations demonstrate a more substantial Dice score improvement of 
%$149.39\%$ 
$0.245$ (from $0.164$ to $0.409$)
in the pre-contrast test domain. Despite being close to the post-contrast distribution as outlined above (e.g., see FID$_{RAD}$ of $0.0385$ between synthetic and real post-contrast test data in Table \ref{tab:image_quality}), the synthetic post-contrast data nonetheless provides relevant pre-contrast signals that allow the post-contrast segmentation model to better generalise to pre-contrast test data. Interestingly, training on only synthetic images without real post-contrast counterparts further improves tumour segmentation performance by $0.077$ (i.e., from $0.409$ to $0.486$) in the post-contrast domain. 

\section{DISCUSSION}
\label{sec:discussion}

Although our results allude to real pre-contrast and synthetic post-contrast proximity calling for further %similarity metrics 
investigation %between 
of the two distributions; they pave the way for future work where already existing and available post-contrast series can be used with synthetic ones from new patients to be evaluated in the pre-contrast domain, without needing to acquire the DCE series. 
Improved model generalizability across domains can be a desirable feature in DCE imaging. For instance, for specific patients only pre-contrast imaging is available, as contrast media injection is to be avoided due to the risk of allergic reactions, missing patient consent, pregnancy or compromise of kidney functions. For other patients, images from some fat-saturated post-contrast phases can closely resemble their pre-contrast equivalents due to low dosage of contrast media or rapid washout effects, requiring models to work reasonably well in both domains.

While the present study explores synthetic post-contrast generation via 2D slice-based synthesis, there is further potential in extending our approach to incorporate the synthesis of 3D volumes. The latter can allow generative models to capture a more comprehensive view of tumour characteristics across all planes. Similarly, the integration of additional imaging modalities (e.g, subtraction imaging, DWI) into both, the generative model as well as the downstream task networks, can provide additional insights on robustness and performance across downstream tasks in different clinical settings, and data availability scenarios.
The reliance on whole-image quality metrics presents a limitation in the context of tasks focusing solely on the tumour region such as radiomics-based tumour treatment response prediction \cite{ground_truth}. By integrating tumour region-specific image quality metrics into evaluation frameworks, such as the herein presented Scaled Aggregate Measure (SAMe), the utility and quality of synthetic contrast injection can be complementarily analysed and correlated with clinical downstream task performance. Similarly, extending 2D-based image quality metrics to 3D can provide additional insights on quality and usefulness of synthetic DCE-MRI.
%Including a more extensive training dataset likely further increases the capabilities of the generative model. 

By iteratively incorporating a higher proportion of tumour-containing slices during training, future work can guide generative models towards capturing more nuanced characteristics of tumour regions (e.g., necrotic tumour cores) to further improve generation quality and downstream segmentation results. Also, acquiring more than one segmentation mask per case will allow to evaluate more clinical dimensions such as bilateral or multifocal cases.
%Exploring additional deep generative models, such as denoising probabilistic diffusion models\cite{sohl2015deep, song2019generative, ho2020denoising}, holds further promise for enhancing the translation of contrast-enhanced features into synthetic images to generate highly realistic and clinically relevant synthetic data \cite{osuala2022data}.
%
We motivate subsequent studies on deep generative DCE-MRI models to analyse and improve upon \emph{baseline 1} from Table \ref{tab:experiments_1_2} not only in the pre-contrast test domain, but potentially also in a \textit{synthetic} post-contrast test domain. In the latter, synthetic contrast can simplify the tumour segmentation task as an alternative to evaluating on real post-contrast cases requiring invasive intravenous contrast injection. Bearing in mind that synthetic test data could contain false-positive tumour hallucinations, such data can nevertheless be a useful tool to localise and highlight potential anomalous regions-of-interest in the MRI volume that can be flagged for further clinical analysis and verification.

In conclusion, our study provides valuable contributions to DCE-MRI synthesis and demonstrates its beneficial application to breast tumour volume segmentation. We further propose SAMe to unify approaches of synthetic data quality assessment and generative model training checkpoint selection.
We identify and discuss limitations and suggest promising future directions to refine our approach and further contribute to improved breast cancer diagnosis and treatment outcomes, ultimately benefiting patients and the medical community.

%%%%%%%%%%%%%%%%%%%  Acknowledgments
\section*{ACKNOWLEDGMENTS}
This project has received funding from the European Union’s Horizon Europe and Horizon 2020 research and innovation programme under grant agreement No 101057699 (RadioVal) and No 952103 (EuCanImage), respectively. Also, this work was partially supported by the project FUTURE-ES (PID2021-126724OB-I00) from the Ministry of Science and Innovation of Spain. We would like to thank Dr Marco Caballo (Radboud University Medical Center, The Netherlands) for providing 254 segmentation masks for the Duke Dataset. We would like to thank Dr Julia Schnabel (Helmholtz Center Munich, Technical University Munich, Germany) for valuable discussions and feedback on this work.

% References
\bibliography{report} % bibliography data in report.bib
\bibliographystyle{spiebib} % makes bibtex use spiebib.bst

\end{document}